\documentstyle[aps,prb,epsfig,multicol]{revtex}
\begin{document}
\title{The mechanism of thickness selection in the Sadler-Gilmer model of polymer crystallization}
\author{Jonathan P.~K.~Doye$^{1,2}$ and Daan Frenkel$^2$}
\address{$^1$ University Chemical Laboratory, Lensfield Road, Cambridge CB2 1EW, UK} 
\address{$^2$ FOM Institute for Atomic and Molecular Physics, Kruislaan 407,\\ 
1098 SJ Amsterdam, The Netherlands}
\date{\today}
\maketitle
\begin{abstract}
Recent work on the mechanism of polymer crystallization
has led to a proposal for the mechanism of thickness selection which differs from 
those proposed by the surface nucleation theory of Lauritzen and Hoffman
and the entropic barrier model of Sadler and Gilmer. 
This has motivated us to reexamine the model used by Sadler and Gilmer.
We again find a fixed-point attractor which describes the dynamical convergence 
of the crystal thickness to a value just larger than the minimum 
stable thickness, $l_{\rm min}$.
This convergence arises from the combined effect of two constraints on the
length of stems in a layer: it is unfavourable for a stem to be shorter than 
$l_{\rm min}$ and for a stem to overhang the edge of the previous layer.
The relationship between this new mechanism and the explanation given by Sadler
and Gilmer in terms of an entropic barrier is discussed.
We also examine the behaviour of the Sadler-Gilmer model when an
energetic contribution from chain folds is included. 
\end{abstract}

\begin{multicols}{2}
\section{Introduction}

On crystallization from solution and the melt many polymers form lamellae
where the polymer chain traverses the thin dimension of the crystal many times folding back
on itself at each surface.\cite{Keller57a,Keller96a}
Although lamellar crystals were first observed over forty years ago
their physical origin is still controversial.
It is agreed that the kinetics of crystallization are crucial since extended-chain crystals are
thermodynamically more stable than lamellae.
However, the explanations for the dependence of the lamellar thickness on temperature
offered by the two dominant theoretical approaches appear irreconcilable.\cite{theorynoteb,alttheory}
The lamellar thickness is always slightly greater than $l_{\rm min}$, the minimum thickness for which
the crystal is thermodynamically more stable than the melt;
$l_{\rm min}$ is approximately inversely proportional to the degree of supercooling.\cite{Barham85}

The first theory, which was formulated by Lauritzen and Hoffman soon after the initial
discovery of the chain-folded crystals,\cite{Lauritzen60,Hoffman76a,Hoffman97}
invokes {\em surface nucleation} of a new layer on the thin side faces of the lamellae as the key process.
It assumes that there is an ensemble of crystals of different thickness,
each of which grows with constant thickness.
The crystals which grow most rapidly dominate this ensemble,
and so the average value of the thickness in the ensemble, which is equated with the observed thickness,
is close to the thickness for which the crystals have the maximum growth rate.
The growth rates are derived by assuming that a new crystalline layer grows by the deposition
of a succession of stems (straight portions of the polymer that traverse the crystal once)
along the growth face.
The two main factors that determine the growth rate are the thermodynamic driving force
and the free energy barrier to deposition of the first stem in a layer.
The former only favours crystallization when the thickness is greater than $l_{\rm min}$.
The latter increases with the thickness of the crystal because of the free energetic cost
of creating the two new lateral surfaces on either side of the stem and makes
crystallization of thicker crystals increasingly slow.
Therefore, the growth rate passes through a maximum at an intermediate value of the thickness
which is slightly greater than $l_{\rm min}$.

The second approach, which was developed by Sadler and Gilmer and has been termed the {\em entropic barrier} model,
is based upon the interpretation of kinetic Monte Carlo simulations \cite{Sadler84a,Spinner95}
and rate-theory calculations\cite{Sadler86a,Sadler87d,Sadler88a}
of a simplified model of polymer crystal growth. 
The model has since been applied to the crystallization of long-chain paraffins,\cite{Sadler87c}
copolymer crystallization\cite{Goldbeck90a} and non-isothermal crystallization.\cite{Goldbeck93a}
As with the surface nucleation approach,
the observed thickness is suggested to result from the competition between a driving force
and a free energy barrier contribution to the growth rate.
However, a different cause for the free energy barrier is postulated.
As the polymer surface in the model can be rough,
it is concluded that the details of surface nucleation of new layers are not important.
Instead, the outer layer of the crystal is found to be thinner than in the bulk;
this rounded crystal profile prevents further crystallization.\cite{Sadler88a}
Therefore, growth of a new layer can only begin once a fluctuation occurs to an entropically
unlikely configuration in which the crystal profile is `squared-off'.
As this fluctuation becomes more unlikely with increasing crystal thickness,
the entropic barrier to crystallization increases with thickness.

Recently, we performed a numerical study of a simple model of polymer 
crystallization.\cite{Doye98c,Doye98b,Doye98d}
Our results suggest that some of the assumptions of the Lauritzen-Hoffman (LH) theory do not 
hold and led us to propose a new picture of the mechanism by which the thickness of 
polymer crystals is determined.
Firstly, we examined the free energy profile for the crystallization of a polymer
on a surface. This profile differed from that assumed by the LH theory
because the initial nucleus was not a single stem but two stems connected by a fold
which grew simultaneously.\cite{Doye98c}

Secondly, 
we performed kinetic Monte Carlo simulations on a model where, similar to the LH approach, 
new crystalline layers grow by the successive deposition of stems across the growth face,
but where some of the constraints present in the LH theory are removed---stems could be 
of any length and grew by the deposition of individual polymer units.
We, therefore, refer to this model as the unconstrained Lauritzen-Hoffman (ULH) model.
Attempts had previously been made to study similar models but at a time when
computational resources were limited, so that only approximate and restricted 
treatments were possible.\cite{Frank60,Lauritzen67b,Point79a,Point79b,DiMarzio82a}

These kinetic Monte Carlo simulations confirmed 
that the initial nucleus was not a single stem.
We also found that
the average stem length in a layer is not determined by the properties of the initial nucleus,
and that the thickness of a layer is, in general, not the same as that of the previous layer.\cite{Doye98b,Doye98d}
Furthermore, during growth lamellar crystals select 
the value of the thickness, $l^{**}$, for which growth with constant thickness can occur,
and not the thickness for which crystals grow most rapidly 
if constrained to grow at constant thickness.
This thickness selection mechanism can be understood by considering
the free energetic costs of the polymer extending beyond the edges of
the previous crystalline layer and of a stem being shorter than $l_{\rm min}$,
which provide upper and lower constraints on the length of stems in a new layer.
Their combined effect is to cause the crystal thickness to converge dynamically 
to the value $l^{**}$ as new layers are added to the crystal. 
At $l^{**}$, which is slightly larger than $l_{\rm min}$, growth with constant thickness then occurs.

Some similarities were found between the behaviour of our model and that of the Sadler-Gilmer (SG) model.
For example, at low supercoolings rounding of the crystal growth front was found to inhibit growth.
However, the mechanism of thickness selection that we found appears, at first sight, to be
incompatible with Sadler and Gilmer's interpretation in terms of an entropy barrier.
Here, we reexamine the SG model in order to determine whether the fixed-point attractors
that we found at large supercooling also occur in the SG model, and 
in order to understand the relationship between the two viewpoints.

\section{Methods}

In the SG model the growth of a polymer crystal 
results from the attachment and detachment of polymer units at the growth face. 
The rules that govern the sites at which these processes can occur 
are designed to mimic the effects of the chain connectivity.
In the original three-dimensional version of the model,
kinetic Monte Carlo simulations were performed to obtain many realizations of the polymer
crystals that result. 
Averages were then taken over these configurations to get the properties of the model.\cite{Sadler84a}
Under many conditions the growth face is rough and the correlations between stems in
the direction parallel to the growth face are weak. 
Therefore, an even simpler two-dimensional version of the model was developed in which 
lateral correlations are neglected entirely, 
and only a slice through the polymer crystal perpendicular to the growth face is considered.\cite{Sadler86a,Sadler88a}
One advantage of this model is that it can be formulated in terms of a set of 
rate equations which can easily be solved.
The behaviour of this new model was found to be very similar to the original three-dimensional
model.

\begin{center}
\begin{figure}
\epsfig{figure=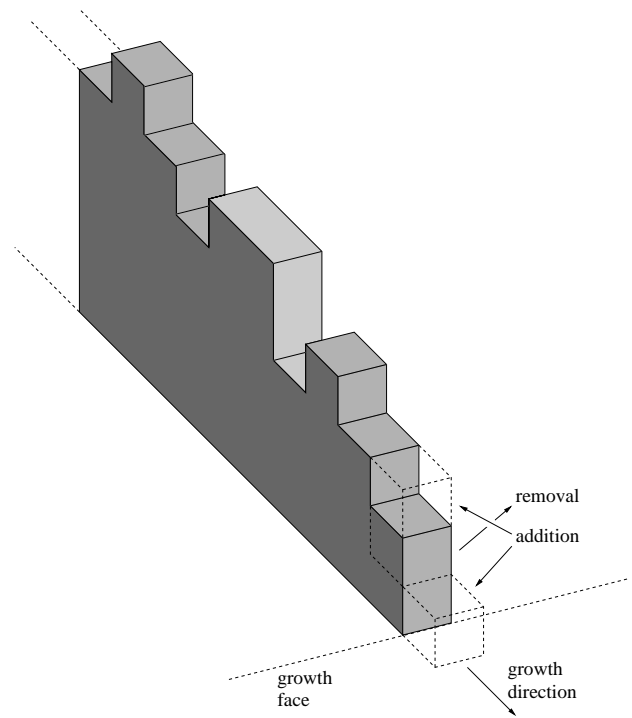,width=8.2cm}
\vglue 1mm
\begin{minipage}{8.5cm}
\caption{\label{fig:moves} A schematic picture of a two-dimensional slice (perpendicular to the
growth face) through a lamellar polymer crystal which forms the basis of the two-dimensional
version of the Sadler-Gilmer model. 
The three possible changes in configuration allowed by the model are shown (the dashed lines
represent the outline of the possible new configurations).
}
\end{minipage}
\end{figure}
\end{center}

The geometry of the model is shown in Figure \ref{fig:moves}.
Changes in configuration can only occur at the outermost stem and stems behind
the growth face are `pinned' because of the chain connectivity.
There are three ways that a polymer unit can be added to or removed from the crystal:
(1) The outermost stem can increase in length upwards. (2) A new stem can be initiated
at the base of the previous stem.
(3) A polymer unit can be removed from the top of the outermost stem.

The set of rate equations that describes the model can be most easily formulated
in a frame of reference that is fixed at the growth face. 
The position of each stem (which is representative of a layer in the three-dimensional crystal)
is then denoted by its distance from the growth face (the outermost stem is at position 1). 
The rate equations are formulated in terms of three quantities:
$C_n(i)$, the probability that the stem at position $n$ has length $i$;
$P_n(i,j)$, the probability that the stem at position $n$ has length $i$ 
and the stem at position $n+1$ has length $j$; and
$f_n(i,j)$, the conditional probability that the stem at position $n+1$ has length $j$
given that the stem at position $n$ has length $i$.
These quantities are related to each other by 
\begin{equation}
C_n(i)=\sum_j P_n(i,j) 
\end{equation}
and 
\begin{equation}
f_n(i,j)=P_n(i,j)/C_n(i).
\end{equation}

The evolution of the system is then described as follows: for $i>1$
\end{multicols}
\begin{equation}
{dP_1(i,j)\over dt}=k^+P_1(i-1,j)+k^-(i+1,j)P_1(i+1,j)+k^-(1,i)P_1(1,i)f_2(i,j)-
                    2 k^+P_1(i,j)-k^-(i,j)P_1(i,j), 
\label{eq:ratea}
\end{equation}
and for $i=1$
\begin{equation}
{dP_1(1,j)\over dt}=k^+C_1(j)+k^-(2,j)P_1(2,j)+k^-(1,1)P_1(1,1)f_2(1,j)-
                    2 k^+P_1(1,j)-k^-(1,j)P_1(1,j),
\label{eq:rateb}
\end{equation}
\begin{multicols}{2}
\noindent where $k^+$ and $k^-(i,j)$ are the rate constants for attachment or detachment of a polymer
unit, respectively.

The first three terms of Equation (\ref{eq:ratea}) correspond to the flux into state $(i,j)$
by extension of the outermost stem, reduction of the outermost stem and the removal
of an outermost stem of length 1, respectively.
The fourth term is the flux out of $(i,j)$ due to the extension of the outermost stem or 
the initiation of a new stem, and the fifth term is the flux out due to a reduction of 
the outermost stem.
The only difference for $i=1$ is that the first term now corresponds to flux into state
$(1,j)$ by initiation of a new stem.

Sadler and Gilmer found $f_n$ to be independent of $n$.\cite{Sadler86a} 
Therefore, $f_2$ can be replaced by $f_1$ in Equations (\ref{eq:ratea}) and (\ref{eq:rateb})
and so the steady-state solution ($dP(i,j)/dt=0$) 
of the above $c_{\rm max}\times c_{\rm max}$ equations can easily be found 
($c_{\rm max}$ is the size of the box in the c-direction).
We used a fourth-order Runge-Kutta scheme\cite{Recipes} to integrate 
the rate equations to convergence starting from 
an initial configuration that, for convenience, we chose to be a crystal of constant thickness.
From $P_1$ it is simple to calculate all $P_n$ by iterative application of the
equations 
\begin{equation}
C_{n+1}(j)=\sum_i P_n(i,j) 
\end{equation}
and 
\begin{equation}
P_{n+1}(i,j)=f(i,j) C_{n+1}(i).
\end{equation}
Far enough into the crystal the properties converge, i.e. $C_{n+1}=C_n$. 
We denote the quantities in this limit as $C_{\rm bulk}$ and $P_{\rm bulk}$.

The rate constants, $k^+$ and $k^-(i,j)$ are related to the 
thermodynamics of the model through 
\begin{equation}
k^+/k^-=\exp(-\Delta F/kT), 
\label{eq:kratio}
\end{equation}
where $\Delta F$ is the change in free energy on addition of 
a particular polymer unit.
The above equation only defines the relative rates and not how the 
the free energy change is apportioned between the 
forward and backward rate constants.
We follow Sadler and Gilmer and choose $k^+$ to be constant.
We use $1/k^+$ as our unit of time.

In the model the energy of interaction between two adjacent crystal units
is -$\epsilon$ and the change in entropy on melting of the crystal is given
by $\Delta S=\Delta H/T_m=2\epsilon/T_m$, where $T_m$ is the melting temperature 
(of an infinitely thick crystal) and 
$\Delta H$ is the change in enthalpy.
It is assumed that $\Delta S$ is independent of temperature.
We can also include the contribution of chain folds to the thermodynamics 
by associating the energy of a fold, $\epsilon_f$, with the initiation of a new stem.\cite{Goldbeck94a} 
Sadler and Gilmer ignored this contribution and so for the sake of comparison 
most of our results are also for $\epsilon_f=0$.

From the above considerations it follows that the rate constants are given by
\begin{eqnarray}
\label{eq:kratio1}
k^-(i,j)&=&k^+ \exp(2\epsilon/kT_m - \epsilon/kT +\epsilon_f/kT) \quad i=1 \\
\label{eq:kratio2}
k^-(i,j)&=&k^+ \exp(2\epsilon/kT_m - 2\epsilon/kT) \quad\quad i\le j, i\ne1 \\
k^-(i,j)&=&k^+ \exp(2\epsilon/kT_m - \epsilon/kT)  \qquad\quad\quad i>j.
\end{eqnarray}
The first term in the exponents is due to the gain in entropy as a result of the removal of a unit 
from the crystal; the second term is due to the loss of contacts between the 
removed unit and the rest of the crystal and the third term (only for $i=1$) is due
to the reduction of the area of the fold surface of the crystal.

Quantities such as the growth rate, $G$,  and the average thickness of the $n^{\rm th}$ layer, 
$\bar{l}_n$, can be easily calculated from the steady-state solution of $P_1(i,j)$:
\begin{equation}
G=k^+ - k^- C_1(1),
\end{equation}
where $k^-$ is given by Equation (\ref{eq:kratio1}),
and 
\begin{equation}
\label{eq:lave}
\bar{l}_n=\sum_i i C_{\rm n}(i).
\end{equation}

We can also estimate the minimum thermodynamically stable thickness, $l_{\rm min}$.
If we assume the crystal to be of constant thickness:
\begin{equation}
\label{eq:lmin}
l_{\rm min}={(\epsilon_f+\epsilon) T_m \over 2\epsilon\Delta T},
\end{equation}
where $\Delta T=T-T_m$ is the supercooling.
However, this value is a lower bound since the roughness in the 
fold surface reduces the surface free energy; 
the resulting surface entropy more than compensates for the increase in surface energy.

As well as solving the rate equations of this model, we sometimes use kinetic Monte Carlo.
This method is more appropriate when examining the evolution of the system towards the steady state.\cite{GGWrate}
At each step in the kinetic Monte Carlo simulation we randomly choose a state, $b$, from the three states
connected to the current state, $a$, with a probability given by
\begin{equation}
P_{ab}={k_{ab}\over\sum_{b'}k_{ab'}},
\end{equation}
and update the time by an increment
\begin{equation}
\Delta t= -{\log\rho\over\sum_{b}k_{ab}},
\end{equation}
where $\rho$ is a random number in the range [0,1].

\section{Results}
\subsection{Mechanism of thickness selection}

The current model has three variables: $kT_m/\epsilon$, $T/T_m$ and $\epsilon_f/\epsilon$.
In this section we only discuss results for $\epsilon_f=0$ as we wish to reexamine 
the same model as that used by Sadler and Gilmer.
Throughout the paper we chose to use $kT_m/\epsilon=0.5$. 
Sadler and Gilmer found that this parameter did not affect the qualitative 
behaviour of the model.\cite{Sadler88a}

\begin{center}
\begin{figure}
\epsfig{figure=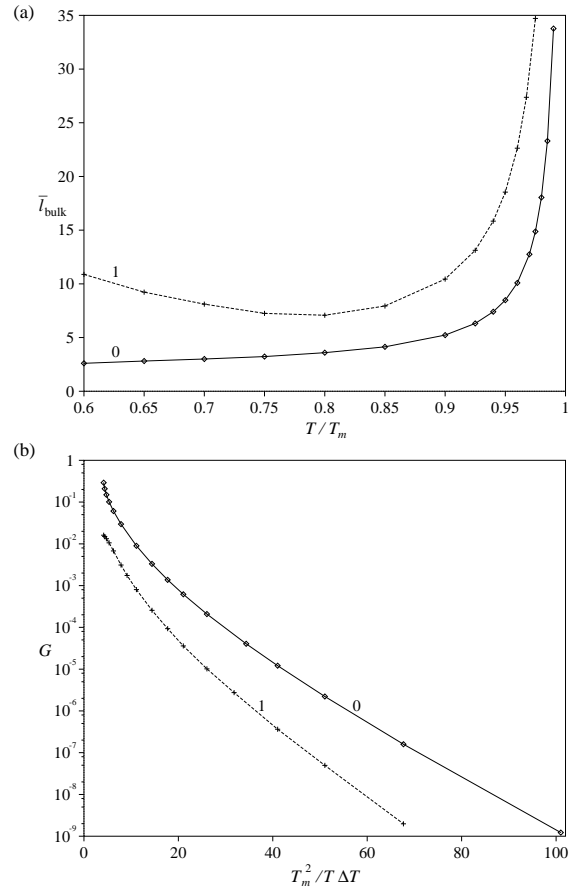,width=7.8cm}
\vglue 1mm
\begin{minipage}{8.5cm}
\caption{\label{fig:l+G} The temperature dependence of the 
(a) crystal thickness and (b) the growth rate. 
The lines are labelled by the value of $\epsilon_f/\epsilon$.
}
\end{minipage}
\end{figure}
\end{center}

\begin{center}
\begin{figure}
\epsfig{figure=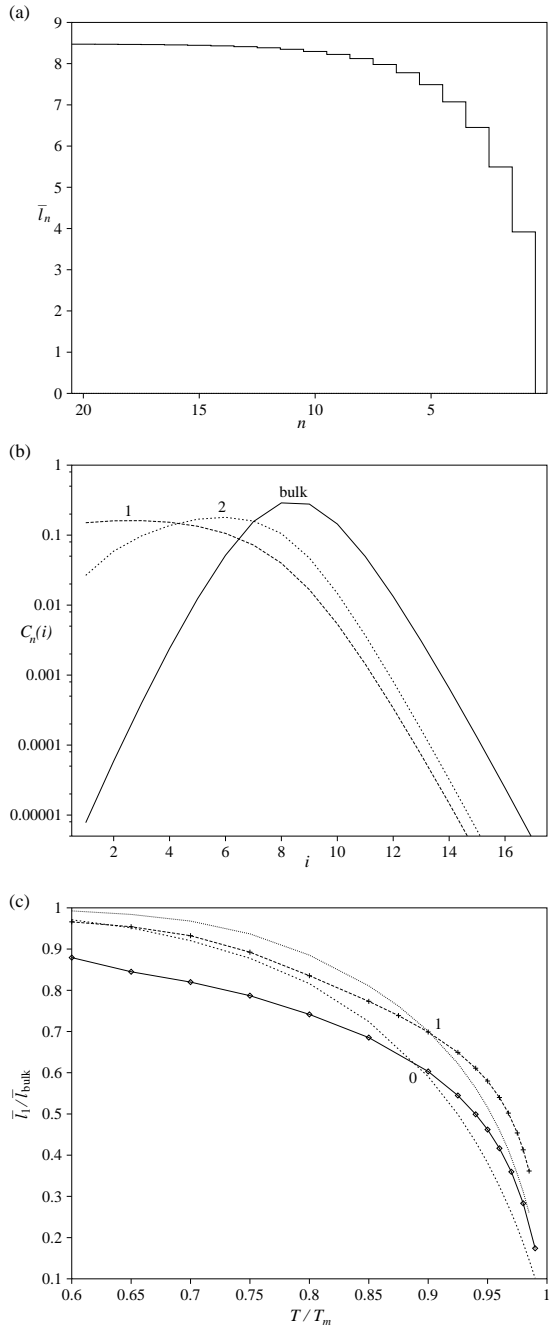,width=7.7cm}
\vglue 1mm
\begin{minipage}{8.5cm}
\caption{\label{fig:profile}(a) The thickness profile of the crystal and 
(b) the probability distribution of the stem length in layer $n$ at $T=0.95T_m$.
The lines are labelled by the value of $n$.
(c) The temperature dependence of the rounding of the crystal profile as measured by
$\bar{l}_{1}/\bar{l}_{\rm bulk}$.
The lines are labelled by the value of $\epsilon_f/\epsilon$,
and those without data points are derived from Equation (\ref{eq:Cl.JD}).
}
\end{minipage}
\end{figure}
\end{center}

Figures \ref{fig:l+G} and \ref{fig:profile} show the typical behaviour of the model.
The crystal thickness, $\bar{l}_{\rm bulk}$, is a monotonically increasing function of the temperature
which goes to $\infty$ at $T_m$. 
The approximately linear character of Figure \ref{fig:l+G}b is in agreement with 
the expected temperature dependence of the growth rate: $G\propto\exp{K_G/T\Delta T}$,
where $K_G$ is a constant.\cite{Keller96a,Gnote} 
The rounding of the edge of the lamellar crystal is clear from the
typical crystal profile shown in Figure \ref{fig:profile}a.
This rounding increases with temperature (Figure \ref{fig:profile}c) suggesting
that it is associated with the greater entropy of configurations with a rounded profile.
The rounding also increases with an increasing value of $kT_m/\epsilon$.\cite{Sadler88a}

The probability distributions of the stem length, $C_n$, of course, 
reflect this rounding (Figure \ref{fig:profile}b). 
The main difference between $C_n$ for different values of $n$ occurs at small stem lengths.
All the $C_n$ seem to have the same approximately exponential decay 
at large stem lengths.
However, the bulk can tolerate few stems shorter than $l_{\rm min}$
for reasons of thermodynamic stability. 
Near the growth face this thermodynamic constraint is more relaxed. 
Moreover, at the growth face there is the kinetic effect that
each new stem must initially be only 
one unit long, thus contributing to $C_1(1)$. 
It is also interesting to note that $C_{\rm bulk}$ is 
approximately symmetrical about its maximum.

The rounding has a major effect on the mechanism of growth.
Since the majority of the short stems at the surface cannot be incorporated
into the bulk, rounding inhibits growth. 
Therefore, a fluctuation in the outer layer to a thickness 
similar to $\bar{l}_{\rm bulk}$ is required before the growth front can advance. 
By this mechanism stems with length less than $l_{\rm min}$ are selected out.
From Figure \ref{fig:profile}a it can be seen that this selection process becomes
more complete as one goes further into the crystal. 

One of the main aims of this paper is to discover whether the 
mechanism of thickness determination that we found for the ULH model\cite{Doye98b,Doye98d} 
also holds for the SG model. 
Central to this mechanism was a fixed-point attractor that related the thickness 
of a layer to the thickness of the previous layer.
As layers in that model once grown could not thereafter change their thickness, the relationship
between the thickness of a layer and the previous layer was independent of the position of the layer in the crystal.
The same is not true of the SG model since there is a zone near to the growth face
where the layers have not yet achieved their final thickness (Figure \ref{fig:profile}a).
Therefore, it is appropriate to look for the attractor in the relationship between the thickness of layers in
the bulk of the crystal.

To do so we define $f'_n(i,j)$ as the conditional probability that the stem at position $n$ has 
length $i$ given that the stem at position $n+1$ has length $j$.
$f'_n(i,j)$ is simply related to quantities that we know:
\begin{equation}
f'_n(i,j)=P_n(i,j)/C_{n+1}(j).
\end{equation}
From this we can calculate $l_n(j)$, which is the average thickness of layer $n$ given that layer
$n+1$ has thickness $j$, using
\begin{equation}
l_n(j)=\sum_i f'_n(i,j)i.
\end{equation}
The actual thickness of the $n^{\rm th}$ layer is simply related to this function by
\begin{equation}
\bar{l}_n=\sum_j C_{n+1}(j) l_n(j).
\end{equation}

\begin{center}
\begin{figure}
\epsfig{figure=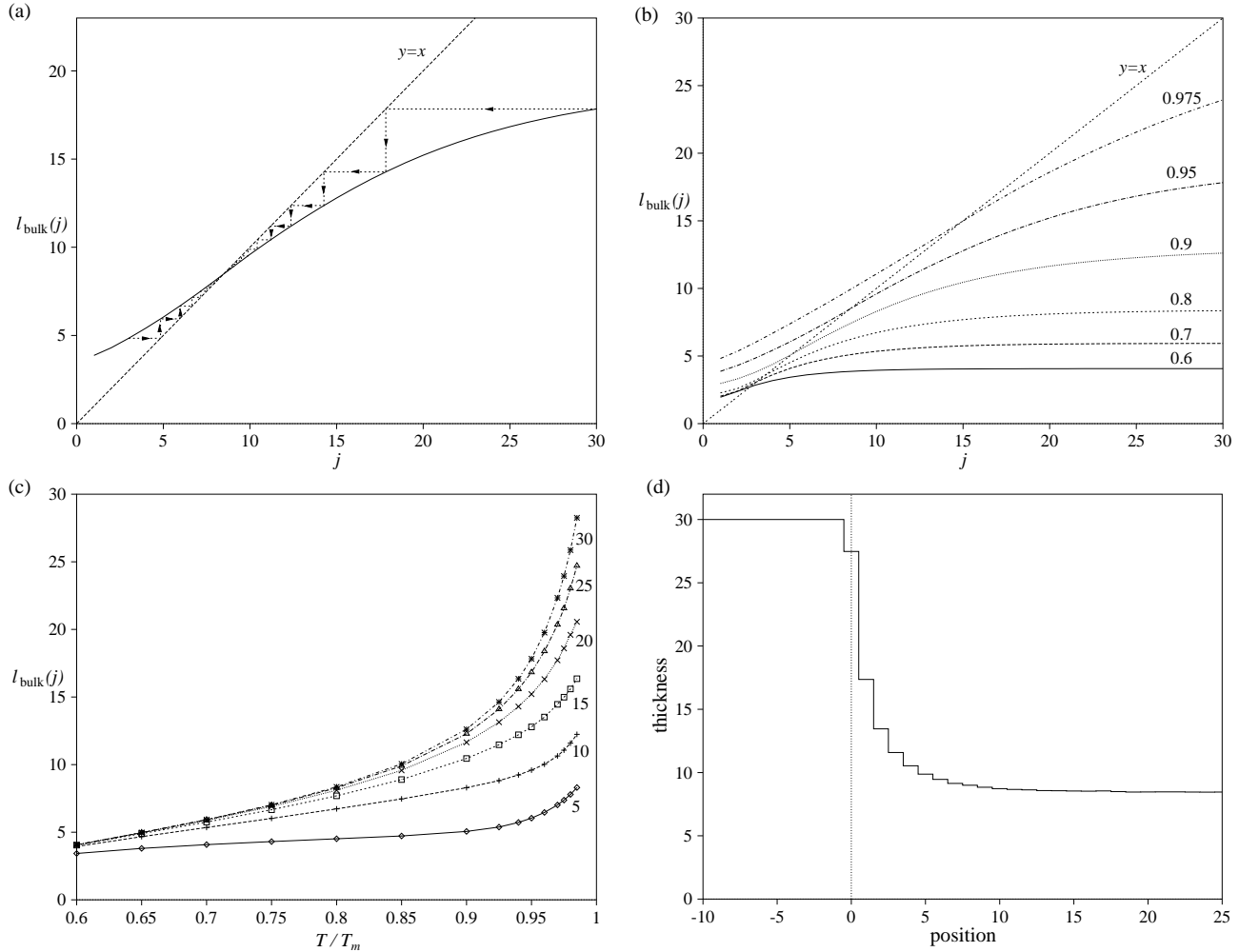,width=17.6cm}
\vglue 1mm
\begin{minipage}{17.6cm}
\caption{\label{fig:attract} 
(a)--(c) The dependence of the thickness of a layer in the bulk of the crystal on the thickness of 
the previous layer, $j$, and the temperature. 
In (a) the fact that the mapping is a fixed-point attractor is illustrated by the dotted arrowed 
lines which show the thickness converging to the fixed point from above and below. $T=0.95T_m$.
In (b) the lines are labelled by $T/T_m$ and in (c) the lines are labelled by the value of $j$.
(d) the average crystal profile that results from an initial crystal that is 30 units thick at
$T=0.95T_m$. The position of the edge of the initial crystal is at zero.
}
\end{minipage}
\end{figure}
\end{center}

Figure \ref{fig:attract}a shows the function $l_{\rm bulk}(j)$ at $T/T_m=0.95$.
As anticipated, the figure has the form of a fixed-point attractor.
The arrowed dotted lines show the thickness of the layer converging to the fixed point
from above and below. 
For example, a layer in the bulk of the crystal which is thirty units thick 
will be followed by one that is 17.8, which will be followed by one that is 14.3, \dots ,
until the thickness at the fixed point, $l^{**}$, is reached.
$l^{**}$ is approximately equal to the crystal thickness, $\bar{l}_{\rm bulk}$.
Figure \ref{fig:attract}d shows the crystal profile that results from starting kinetic Monte 
Carlo simulations from a crystal that is initially 30 units thicks.
The convergence of the thickness to $l^{**}$ is apparent and 
the resulting step on the surface is similar to that produced by a temperature jump.\cite{Bassett62,Dosiere86a} 
Figure \ref{fig:attract}b shows that similar maps with a fixed point occur at all temperatures.
As the temperature increases the slope of the curves become closer to 1 and so the
convergence to $l^{**}$ becomes slower.

\begin{center}
\begin{figure}
\epsfig{figure=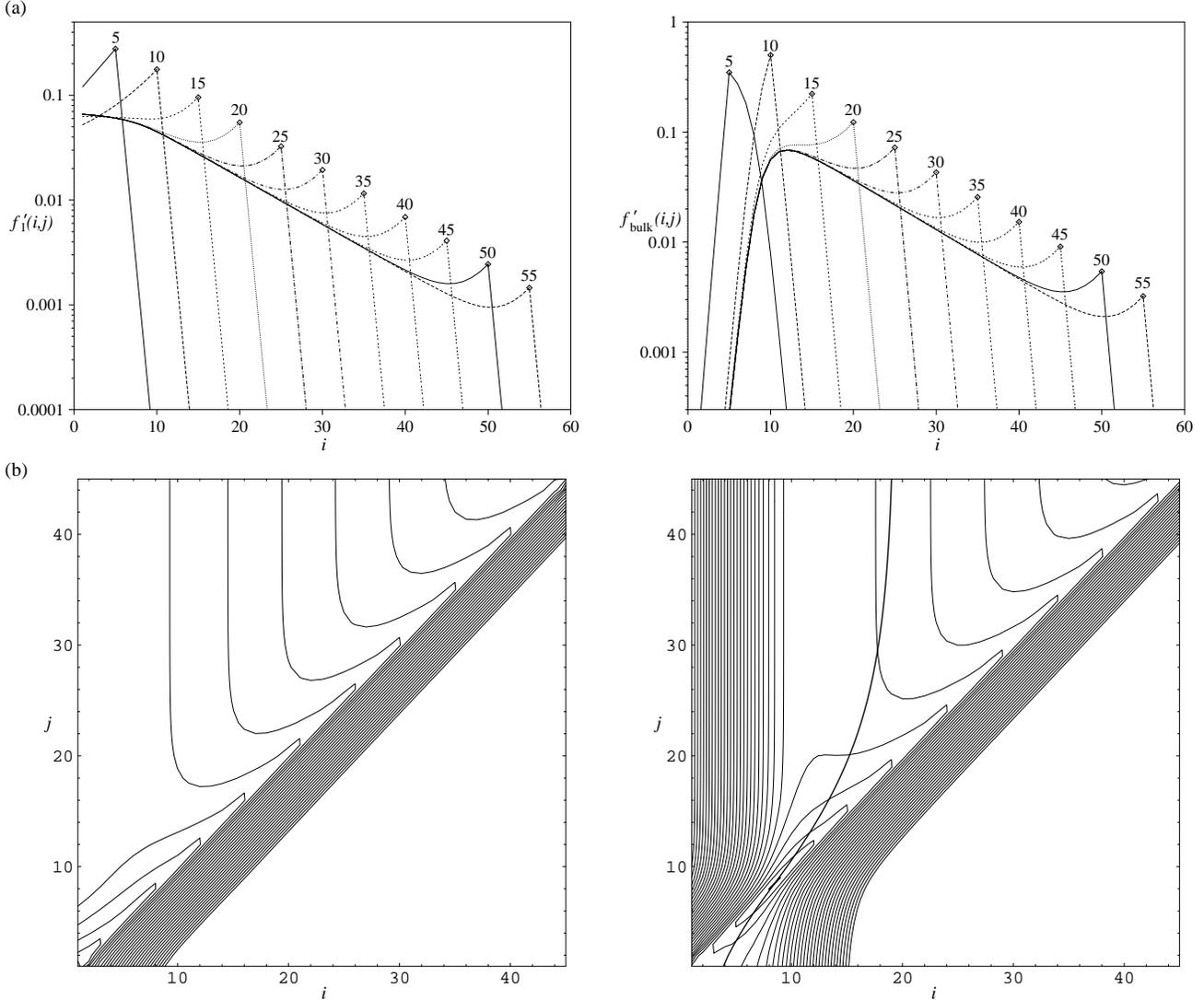,width=17.6cm}
\vglue 1mm
\begin{minipage}{17.6cm}
\caption{\label{fig:pstem} 
(a) Probability distributions of the stem length in layer $n$ given that the layer 
$n+1$ is of thickness $j$ at $T=0.95T_m$.
The labels give the value of $j$ and the data points are for $i=j$.
(b) shows a two-dimensional contour plot of $\log(f'_n(i,j))$ at $T=0.95T_m$.
(The contours have a spacing of 0.5 and are measured with respect to the maximum in the probability. 
For clarity we only include the first 30 contours below the probability maximum;
the probability does actually continue to fall rapidly in the blank bottom right corner.)
The lefthand figures are for the outermost layer and 
the righthand figures for a layer in the bulk.
The thicker line in the righthand panel of (b) displays the fixed-point attractor, $l_{\rm bulk}(j)$.
}
\end{minipage}
\end{figure}
\end{center}

To understand why the function $l_n(j)$ has the form of a fixed-point attractor
we examine probability distributions of the stem length when the previous layer had a specific thickness.
In Figure \ref{fig:pstem} we show $f'_{\rm bulk}(i,j)$ for different $j$.
These probability distributions have a number of features.
Firstly, the probability that the stem length is smaller than $l_{\rm min}$
drops off exponentially. 
This is a simple consequence of the thermodynamics.
Secondly, the probability that the stem length is larger than the thickness of
the previous layer also drops off exponentially. 
This is again related to the thermodynamics. 
The absence of an interaction of the polymer with the surface makes it 
unfavourable for a stem to extend beyond the edge of the previous layer.
This feature also causes a build-up in probability at stem lengths at or just 
shorter than the thickness of the previous layer.
Therefore, as we saw in our previous model,\cite{Doye98b,Doye98d} there is a range of 
stem lengths between $l_{\rm min}$ and the thickness of the previous layer which 
are thermodynamically viable, and it is the combined effect of the two constraints
that causes the thickness to converge to the fixed point.

\end{multicols}
\begin{multicols}{2}

Thirdly, when the previous layer is thick the probability also falls off exponentially with
increasing stem length in the range $l_{\rm min}<i<j$.
This decay of the probability is much less rapid than for stems that 
overhang the previous layer and is a kinetic effect. 
At each step in the growth of a stem there is a certain probability, $p_{\rm new}$, that a new stem 
will be successfully initiated, thus bringing the growth of the original stem to an end.
It therefore follows that the probability will decay with an exponent $\log(1-p_{\rm new})$.
It is for this reason that the thickness of a new layer will remain finite even 
if the previous is infinitely thick. 
Comparison of $f_{\rm bulk}$ at different temperatures
shows that this rate of decay increases with decreasing temperature. 
This is because the thermodynamic driving force is greater at lower temperature making 
successful initiation of a new stem more likely (at higher temperature a new stem is much more likely 
to be removed).

The relationship between the probability distributions and the fixed-point attractor can 
be most clearly seen from the contour plot of $\log(f'_{\rm bulk}(i,j))$ in Figure \ref{fig:pstem}b.
It clearly shows the important roles played by $l_{\rm min}$ and $j$ in constraining the thickness of the next layer.
When the previous layer is larger than $l^{**}$ the average stem length must lie between 
$l_{\rm min}$ and $j$ causing the crystal to become thinner. 
As $j$ approaches $l_{\rm min}$ the probability distributions become narrower and the
difference in probability between the stem length being greater than $j$ and smaller than $j$
decreases until the fixed point $l^{**}$ is reached where the average stem length is the same as the
thickness of the previous layer. This explains why $l^{**}$ is slightly larger than $l_{\rm min}$.
For $j<l^{**}$ the asymmetry of the probability distribution is reversed.
These results confirm that the mechanism for thickness determination in the SG model 
is very similar to that which we found in the ULH model.\cite{Doye98b,Doye98d}

\end{multicols}
\begin{multicols}{2}

We have also shown our results for $f'_1(i,j)$ in Figure \ref{fig:pstem}.
The main difference from $f'_{\rm bulk}$ results from the fact that stems shorter
than $l_{\rm min}$ can be present in the outermost layer. 
The lack of this constraint causes the outer layer to be thinner than the bulk,
and therefore the crystal profile to be rounded.
The two modes of exponential decay at larger stem length that we noted for $f'_{\rm bulk}$ 
are clearly also present in $f'_1$.

\begin{center}
\begin{figure}
\epsfig{figure=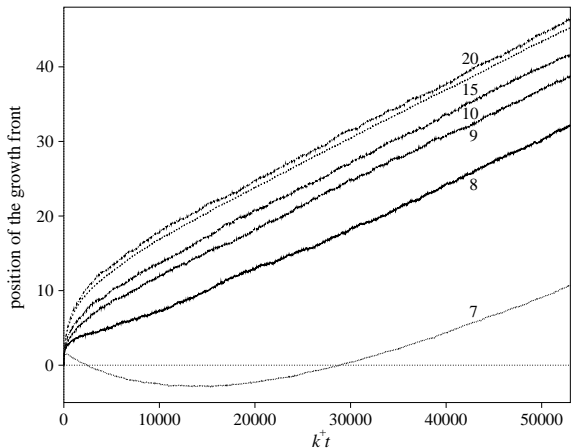,width=8.2cm}
\vglue 1mm
\begin{minipage}{8.5cm} 
\caption{\label{fig:kmc} 
Growth from initial crystals of different thickness at
$T/T_m=0.95$. The initial position of the growth front is defined to be zero. 
The results were obtained from kinetic Monte Carlo simulations.
The lines are labelled by the thickness of the initial crystal.
}
\end{minipage}
\end{figure}
\end{center}

Another finding from our work on the ULH model was that the rate of growth actually slowed down 
as the thickness of a crystal converged to $l^{**}$ from above.\cite{Doye98b,Doye98d}
To determine whether this behaviour also holds for the SG model
we performed kinetic Monte Carlo simulations from initial crystals which had different thicknesses. 
As can be seen from Figure \ref{fig:kmc} the initial growth rate depends on the 
thickness of the initial crystal,
but the final growth rate is independent of this initial thickness (on the right-hand side
of the figure the lines are all parallel) because the thicknesses of all the
crystals have now converged to $l^{**}$. In agreement with our previous results
the initial growth rate increases with the initial crystal thickness.
Interestingly for a crystal with an initial thickness of 7 (less than $l_{\rm min}$)
the crystal initially melts causing the growth face to go backwards.
Only once a fluctuation causes the nucleation of some thicker layers at the growth front
can the crystal begin to grow again. The thinner the crystal the more difficult is this 
nucleation event.

Although we have only presented results for the two-dimensional version of the 
Sadler-Gilmer model in this section, we should note that in the three-dimensional version of 
the model we also found that the thickness of a crystal converged to the same value 
whatever the starting configuration of the crystal.

\subsection{Sadler and Gilmer's entropic barrier}

The description of the mechanism of thickness selection that we give above is 
in terms of the fixed-point attractor, $l_{\rm bulk}(j)$. 
It makes no mention of the growth rate (or a maximum in that quantity) 
in contrast to the LH surface nucleation theory and 
Sadler and Gilmer's own interpretation of the current model.\cite{Sadler86a,Sadler87d,Sadler88a} 
This of course raises the question, ``Which viewpoint is correct?''
In this section we examine Sadler and Gilmer's argument again to determine whether it provides 
a complementary description to the picture presented above or whether the two pictures
are incompatible in some respects. 
This reexamination is complicated by the fact that two slightly different 
arguments were presented.

\begin{center}
\begin{figure}
\epsfig{figure=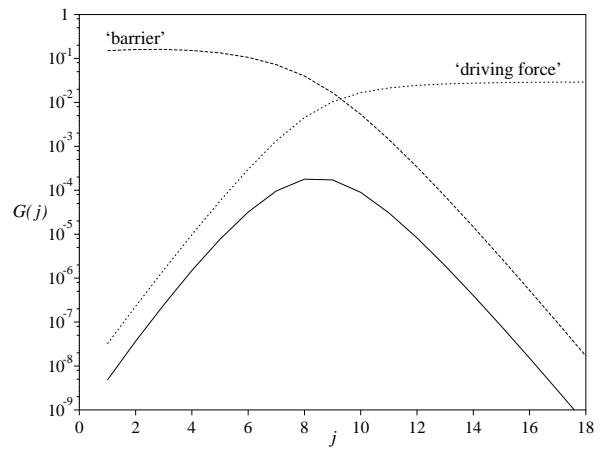,width=8.2cm}
\vglue 1mm
\begin{minipage}{8.5cm}
\caption{\label{fig:G.comp} Decomposition of the growth rate (solid line) into
a barrier and a driving-force term using Equation (\ref{eq:Gjb}) at $T/T_m=0.95$.
}
\end{minipage}
\end{figure}
\end{center}

In the first the growth rate was decomposed into the terms\cite{Sadler88a}
\begin{equation}
G(i)=k^+ C_1(i) - k^-(1,i) P_1(1,i),
\label{eq:Gj}
\end{equation}
which satisfy $G=\sum_i G(i)$.
The function $G(i)$ has a maximum close to $l_{\rm bulk}$ (Figure \ref{fig:G.comp}).
It is expected that 
\begin{equation}
\bar{l}_{\rm bulk}=\sum_i i G(i)/ G.
\label{eq:l+G}
\end{equation}
Sadler and Gilmer argue that the maximum in $G(i)$ causes the crystal thickness to take 
the value $l_{\rm bulk}$.

Firstly, we should note that this explanation is very different from the
maximum growth-rate criterion in the LH theory. 
In the LH theory the growth rate has a maximum with respect to the thickness
in a fictitious ensemble of crystals of different thickness, 
each of which grows with constant thickness.
Rather, $G(i)$ is a characterization of that one steady state
where growth with constant average thickness occurs.

Secondly, as a characterization of the steady state Equation (\ref{eq:l+G}) is correct.
As $G(i)$ is simply the rate of incorporation of stems of length $i$ into the crystal,
at the steady state it should obey the equation:\cite{Goldbeck93a,check} 
\begin{equation}
G(i)=G C_{\rm bulk}(i).
\label{eq:GC}
\end{equation}
Comparison of Figures \ref{fig:profile}b and \ref{fig:G.comp} provides a simple visual 
confirmation of the validity of this equation.
Equation (\ref{eq:l+G}) simply follows from the above equation and Equation (\ref{eq:lave}).
The above equation also implies that explanations in terms of $G(i)$ 
or $C_{\rm bulk}(i)$ should be equivalent.

Thirdly, although the steady state must be stable against fluctuations (the rate
equations would not converge to it if it was not)
$G(i)$ does not in itself explain how convergence to the steady state is achieved 
or how fluctuations to larger or smaller thickness decay away.
$G(i)$ only characterizes the steady state and not the system when it
has been perturbed from the steady state.
For example, the maximum in $G(i)$ does not imply that crystals which are thicker
than $l_{\rm bulk}$ grow more slowly. Rather, as we showed earlier, these crystals
initially grow more rapidly, but this growth causes the thickness to converge
to $l^{**}$.

Sadler and Gilmer then continue by examining the form of $G(i)$. 
Rearranging Equation (\ref{eq:Gj}) leads to\cite{Sadler88a}
\begin{equation}
G(i)=k^+ \left(1 - {k^-(1,i) P_1(1,i)\over k^+ C_1(i)}\right) C_1(i) 
\label{eq:Gjb}
\end{equation}
This equation now has the form of a driving force term 
multiplied by a barrier term.
The barrier term, $C_1(i)$, is the probability that the system is at the transition state
and the driving-force term ($k^+$ multiplied by the term in the brackets) is the
rate at which this transition state is crossed.
The two terms are shown in Figure \ref{fig:G.comp} and it can be seen that their forms 
agree with the above interpretation. 
At small $i$, $G(i)$ is small because of the small driving force and 
at large $i$ the barrier term leads to the decrease in $G(i)$. 
This leads to a maximum in the growth rate at an intermediate value of $i$ when the barrier and driving force terms
have intermediate values.

It should be noted that the term we have labelled the driving force does not have 
exactly the behaviour one might expect. 
In contrast to Sadler and Gilmer's suggestion,\cite{Sadler88a}
this term is not simply proportional to $i-l_{\rm min}$ near to $l_{\rm min}$.
For, as is evident from Equation (\ref{eq:GC}), 
$G(i)$ can never become negative even at $i<l_{\rm min}$.
This is because $G(i)$ is not the rate of growth of a crystal of thickness $i$ but
the rate of incorporation of stems of length $i$ when the system is at the steady state;
there is always a small probability that a stem with length less than
$l_{\rm min}$ is incorporated into the crystal.

The barrier term, $C_1$, has a clear interpretation; 
it quantifies the effect of the rounding of the crystal growth front in inhibiting growth.
The rate-determining step in the growth process is the fluctuation to a more unlikely 
squared-off configuration that has to occur before growth is likely to proceed. 
However, Sadler and Gilmer interpret $C_1$ in terms of what they
call an {\it entropic barrier}. 
Firstly, we note that a decay in $C_1(i)$ does not necessarily 
imply an increase in the free energy with $i$ because $C_1(i)$ is a steady-state probability 
distribution under conditions where the crystal is growing, 
whereas the Landau free energy for the outermost stem, $A_1(i)$, is related to the equilibrium probability 
distribution through 
\begin{equation}
A_1(i)=A-kT\log C_1^{\rm eq}(i),
\label{eq:Landau}
\end{equation}
where $A$ is the Helmholtz free energy.
The outermost layer may not have reached a state of thermodynamic equilibrium.

\begin{center}
\begin{figure}
\epsfig{figure=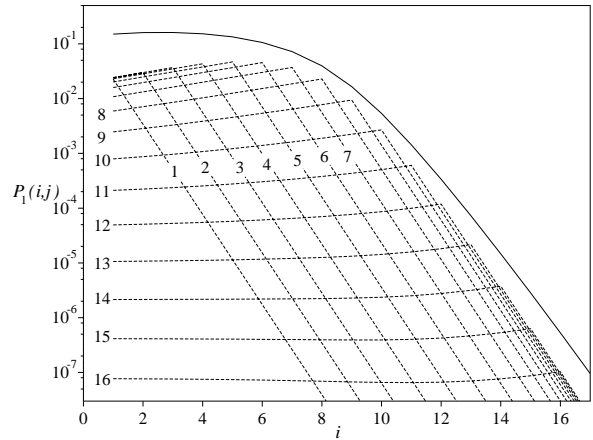,width=8.2cm}
\vglue 1mm
\begin{minipage}{8.5cm}
\caption{\label{fig:pl1} The solid line is $C_1(i)$ and the dashed lines are the 
contributions to $C_1$ from configurations where the thickness of layer 2 is $j$
($P_1(i,j)=C_2(j)f'_1(i,j)$). Each dashed line is labelled by the value of $j$.
$T=0.95T_m$.
}
\end{minipage}
\end{figure}
\end{center}

Secondly, and more importantly, their explanation makes little mention of the 
limiting effect of the thickness of the previous layer. 
However, if we examine the contributions
to $C_1(i)$ from configurations where the thickness of layer 2 is $j$ (Figure \ref{fig:pl1})
it can be clearly be seen that these contributions decay sharply for $i>j$. 
This is because it is unfavourable for a stem to extend beyond the previous layer. 
The combination of this decay in $f'_1(i,j)$ (Figure \ref{fig:pstem}) with the rounding 
inherent in $C_2(j)$ (Figure \ref{fig:profile}b) causes $C_1(i)$ to decay as $i$ increases.

Sadler and Gilmer also considered a second approach to account for the 
thickness selection of lamellar crystals.
In this approach a maximum in $G(l)$, the growth rate of a crystal of thickness $l$, 
is sought.\cite{Sadler87d}
However, by doing so this approach suffers the same
weakness as the LH maximum growth criterion. 
The implicit assumption that a crystal of any thickness continues to grow at that thickness
is contradicted both by the behaviour of the current model (e.g.\ Figure \ref{fig:attract}d)
and by experiment.\cite{Bassett62,Dosiere86a} 

As with the previous approach the growth rate is divided into a barrier and a
driving force term. The barrier term is $C(l)$, the probability that the system had a 
squared-off configuration (i.e.\ $l_n=l$ for all $n$).
Sadler and Gilmer argue that $C(l)$ is an exponentially decreasing function of $l$,
because `it seems likely the number of tapered configurations will go up approximately 
exponentially with $l$'.\cite{Sadler86a}

Later, Sadler quantified this argument by calculating $C(l)$ for the case 
that $l_n\le l_{n+1}$:\cite{Sadler87d}
\begin{equation}
C(l)=\prod_{j=1}^{l-1}{1\over 1+(k^-/k^+)^j}
\end{equation}
However, this result appears to be incorrect. We find:
\begin{equation}
\label{eq:Cl.JD}
C(l)=\prod_{j=1}^{l-1} \left[ 1-(k^-/k^+)^j \right].
\end{equation}
The original and the present derivations are given in the Appendix.\cite{Clmelt}
However, the two expressions do have a similar functional dependence 
on $l$ and temperature,
and so the error does not affect the substance of the argument.

\begin{center}
\begin{figure}
\epsfig{figure=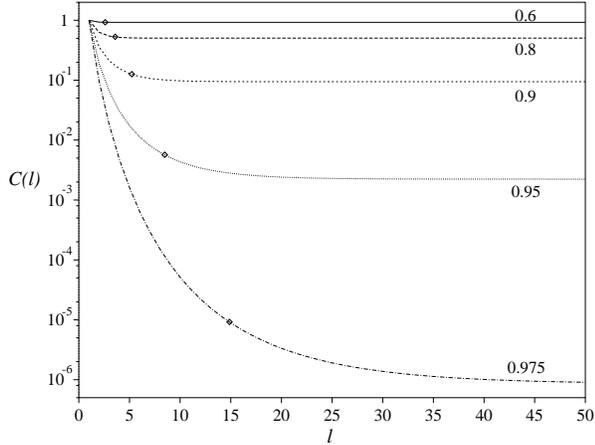,width=8.2cm}
\vglue 1mm
\begin{minipage}{8.5cm}
\caption{\label{fig:psquare} 
The probability of a squared-off configuration, $C(l)$, at different
temperatures as labelled. The diamonds mark the position of $\bar{l}_{\rm bulk}$
at that temperature.
}
\end{minipage}
\end{figure}
\end{center}

In Figure \ref{fig:psquare}a we show the dependence of $C(l)$ on $l$ at a
number of temperatures. It can be seen that $C(l)$ does not decay exponentially, but after 
an initial decay reaches an asymptote at large $l$.\cite{Clrate} 
The origins of this behaviour in Equation (\ref{eq:Cl.JD}) is easy to understand. 
For $T<T_m$, $k^-/k^+<1$ and therefore
\begin{equation}
\lim_{l\to\infty}{P(l+1)\over P(l)}=\lim_{l\to\infty}\left[1 - \left(k^-/k^+\right)^l\right]=1.
\end{equation}
The rate of decay to this asymptote depends on the value of $k^-/k^+$.
As the temperature is decreased, the magnitude of $\Delta F$ increases and therefore
$k^-/k^+$ decreases (Equation \ref{eq:kratio}) and 
$C(l)$ decays to its asymptote more rapidly.
The curves in Figure \ref{fig:psquare} show just this behaviour.

The physical significance of these results is clear. 
For thick crystals a squared-off configuration does not become any more unlikely 
as the thickness is increased.
In particular, the free energy barrier to growth caused by rounding of the crystal 
profile does not grow exponentially with $l$ as Sadler and Gilmer have argued.

It should also be stressed that the rounding of the crystal growth front that is expressed by 
$C(l)$ is not just a result of entropy. 
The number of configurations with an outermost stem of length $i$ is independent of $i$.
Rounding of the crystal profile occurs because those fluctuations that lead to a stem near the 
edge of the crystal being thicker than the previous layer 
are energetically disfavoured compared to those which lead to a stem that is shorter than the previous layer.
Again we see the crucial role played by the free energetic penalty for a stem extending beyond the previous layer. 
Sadler implicitly included this effect in the calculation of $C(l)$ by imposing 
the condition $l_{n}\le l_{n+1}$.
Rounding of the crystal profile results from an interplay of energetic and 
entropic contributions to the free energy.

\begin{center}
\begin{figure}
\epsfig{figure=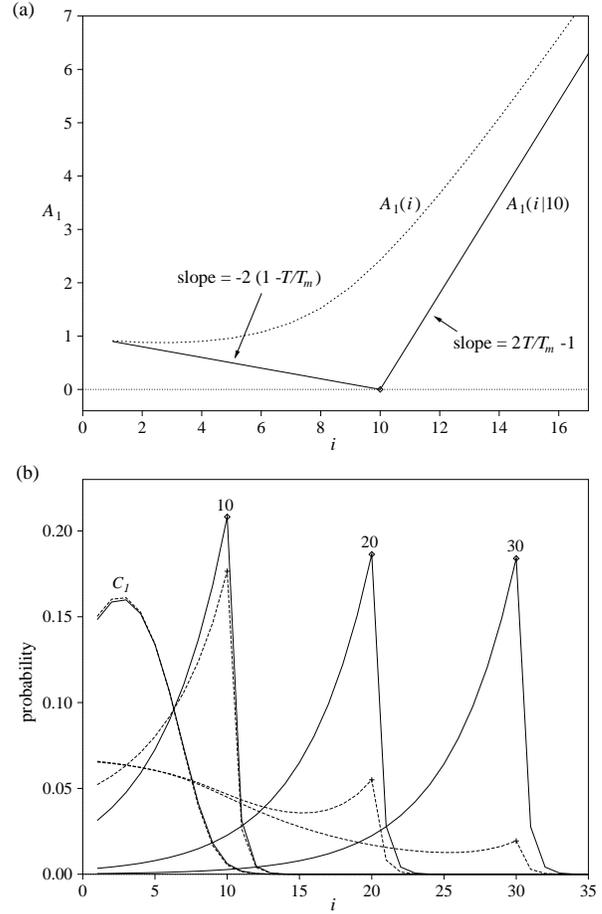,width=8.2cm}
\vglue 1mm
\begin{minipage}{8.5cm}
\caption{\label{fig:A1} 
(a) $A_1(i|j)$ for $j=10$ and $A_1(i)$. 
(b) Comparison of the steady-state (dashed lines) and equilibrium (solid lines) values 
of the probabilities $f'_1(i,j)$ and $C_1(i)$.
The equilibrium values were derived from Equations (\ref{eq:f1eq}) and (\ref{eq:C1eq}).
For $f'_1(i,j)$ the lines are labelled by the value of $j$. $T=0.95 T_m$
}
\end{minipage}
\end{figure}
\end{center}

\subsection{The free energy of the outermost stem}

As the free energy function that is experienced by the outermost stem plays 
such an important role in both the LH and SG mechanisms of thickness determination
here we calculate this quantity for the current model.
This also allows us to examine the magnitude of non-equilibrium effects.

Firstly, $A_1(i|j)$, 
the free energy of the outermost stem when the thickness of layer 2 is $j$, is simply
\begin{eqnarray}
A_1(i|j)/\epsilon &=& 2iT/T_m - 2i+1 \qquad i \le j \\
A_1(i|j)/\epsilon &=& 2iT/T_m - i-j+1 \quad i > j.
\end{eqnarray}
For $i\le j$ the free energy decreases with increasing i because the energy
gained from crystallization is greater than the loss in entropy.
However, for $i>j$ the new polymer units added to the crystal no longer gain
any energy from interactions with the previous layer causing the free energy to increase. 
The example free energy profile, $A_1(i|10)$, shown in Figure \ref{fig:A1}a exhibits these features.

From $A_1(i|j)$ it is easy to calculate the value of $f'_1(i,j)$ if the
outermost stem is at equilibrium:
\begin{equation}
f'^{\rm eq}_1(i,j)={\exp(-A_1(i|j)/kT)\over \sum_{i'} \exp(-A_1(i'|j)/kT)}.
\label{eq:f1eq}
\end{equation}
In Figure \ref{fig:A1}b we show $f'^{\rm eq}_1(i,j)$ at a number of values of $j$.
$f'^{\rm eq}_1(i,j)$ is always peaked at $j$ the position of the free energy minimum
in $A_1(i|j)$ and the asymmetry in the probability distribution (the stem length
is more likely to be less than $j$ than greater than $j$) reflects the 
asymmetry in the free energy profile.
Comparison of the equilibrium and steady-state values of $f'_1(i,j)$
shows that the two are in reasonable agreement for small $j$ but
the discrepancy increases with $j$.
The steady-state values of $f'_1(i,j)$ are always larger at small values of 
$j$ because of the bias introduced by the successful initiation
of a new stem before equilibrium is reached for the outermost stem; 
this kinetic effect was noted in our earlier discussion of the form of $f'(i,j)$.
As $j$ increases it becomes increasingly likely that a new stem is initiated 
before the current stem has reached a length $j$. 

We can also obtain $C_1^{\rm eq}(i)$ from $f'^{\rm eq}_1(i,j)$: 
\begin{equation}
C_1^{\rm eq}(i)=\sum_j f'^{\rm eq}_1(i,j)C_2(j).
\label{eq:C1eq}
\end{equation}
Using the steady-state values of $C_2$ in the above equation we found that
$C_1^{\rm eq}$ and $C_1$ are virtually identical (Figure \ref{fig:A1}b)
reflecting the good agreement between $f'_1(i,j)$ and $f'^{\rm eq}_1(i,j)$ 
for small $j$.

As $C_1^{\rm eq}(i)$ is also related to $A_1(i)$, the free energy of a stem in the outermost layer
of length $i$, through Equation (\ref{eq:Landau}),
it follows that 
\begin{eqnarray}
A_1(i)=&-&kT \log \sum_j {C_2(j) \exp(-A_1(i|j)/kT) \over \sum_{i'} \exp(-A_1(i'|j)/kT)}\nonumber\\
       &+&kT \log \sum_j {C_2(j) \over \sum_{i'} \exp(-A_1(i'|j)/kT)},
\end{eqnarray}
where we have chosen the zero of $A_1(i)$ so that $A_1(1)=A_1(1|j)=2T/T_m-1$.
Figure \ref{fig:A1}a shows that $A_1(i)$ rises as $i$ increases.
There is a free energy barrier that increases with stem length.
As with $C_1$ this increase results from the combination of the
rounding present in $C_2$ and the free energy barrier to 
extension of a stem beyond the edge of the previous layer.

\subsection{Effect of $\epsilon_f$}

Although we found a similar mechanism of thickness selection in the SG model
as in the ULH model,\cite{Doye98b,Doye98d} 
there are a number of differences in the behaviour of the models.
Firstly, the crystal profile in the SG model is always rounded whereas in 
the ULH model rounding does not occur at larger supercoolings.
Secondly, $l_{\rm bulk}(j)$ in the SG model is a monotonically increasing 
function of temperature, whereas it increases at lower temperatures in the ULH model. 
This increase is because the free energy barrier for the formation of a fold (and
hence the initiation of a new stem) becomes increasingly difficult to scale, 
so on average a stem continues to grow for longer. This effect is probably also
partly responsible for the lack of rounding at larger supercoolings.
One possible source of some of the differences between the models is the neglect 
of the energetic cost of forming a fold. In this section we consider the effect of non-zero
$\epsilon_f$ by presenting results for $\epsilon_f=\epsilon$.

\begin{center}
\begin{figure}
\epsfig{figure=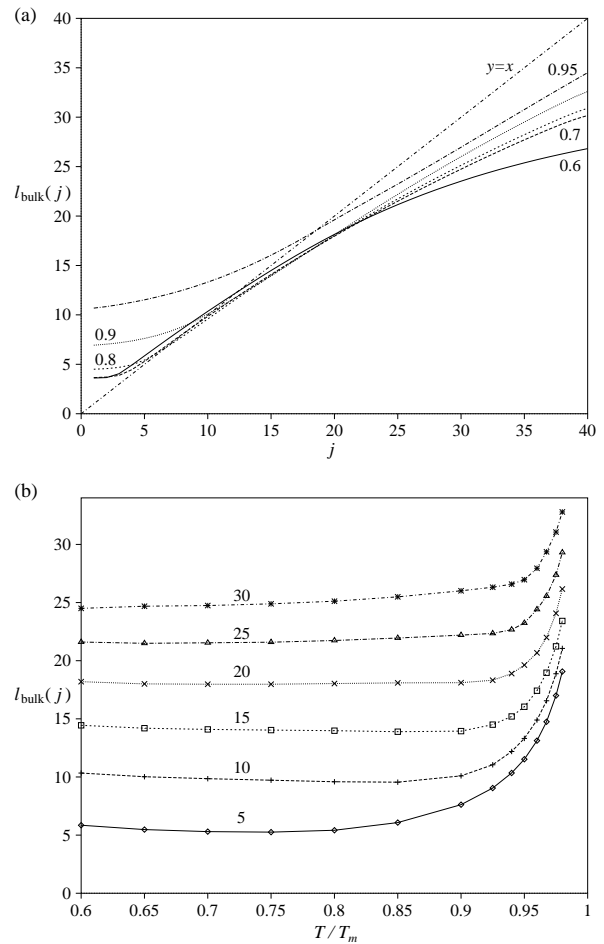,width=8.2cm}
\vglue 1mm
\begin{minipage}{8.5cm}
\caption{\label{fig:attract.ef} 
The dependence of the thickness of a layer in the bulk of the crystal on the thickness of 
the previous layer, $j$, and the temperature when $\epsilon_f=\epsilon$.
In (a) the lines are labelled by $T/T_m$ and in (b) the lines are labelled by the value of $j$.
}
\end{minipage}
\end{figure}
\end{center}

One effect of a non-zero $\epsilon_f$ is to increase $l_{\rm min}$ (Equation \ref{eq:lmin})
and so $\bar{l}_{\rm bulk}$ is always larger than for $\epsilon_f=0$ (Figure \ref{fig:l+G}a).
Furthermore, $\bar{l}_{\rm bulk}$ is now a non-monotonic function of temperature; it
gradually begins to increase below $T=0.8T_m$. We will explain the cause of 
this increase later in this section.
The growth rate is slower than for $\epsilon_f=0$,
because the energetic cost of forming a fold makes initiation of a new stem more difficult.
However, the temperature dependence of $G$ is fairly similar (Figure \ref{fig:l+G}b).

\begin{center}
\begin{figure}
\epsfig{figure=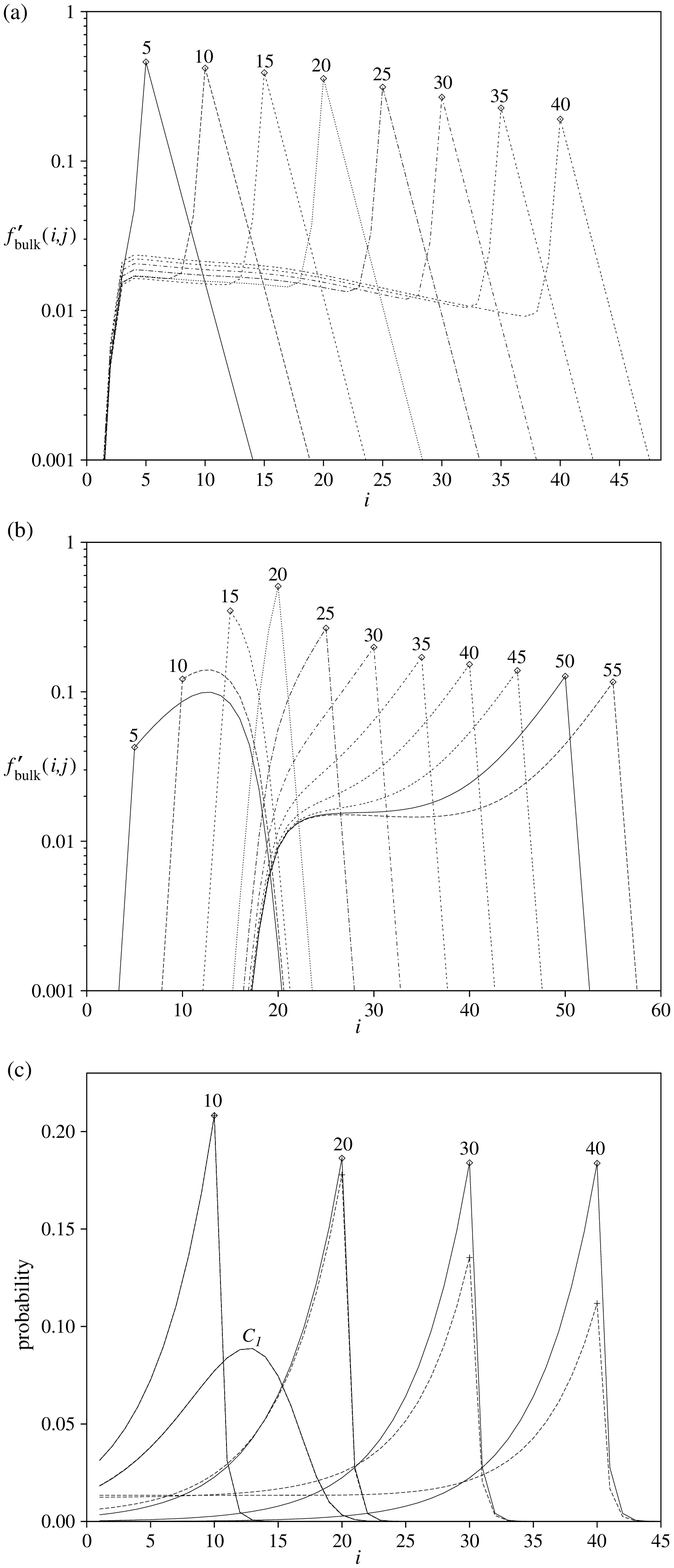,width=8.2cm}
\vglue 1mm
\begin{minipage}{8.5cm}
\caption{\label{fig:pstem.ef} 
Probability distributions of the stem length in a layer in the bulk of the crystal
given that the previous layer is of thickness $j$ at (a) $T=0.6T_m$, and (b) $T=0.95T_m$.
In (a) and (b) the labels give the value of $j$ and the data points are for $i=j$.
(c) Comparison of the steady-state (dashed lines) and equilibrium (solid lines) 
values of the probabilities $C_1(i)$ and $f'_1(i,j)$. 
The equilibrium values were derived from Equations (\ref{eq:f1eq}) and (\ref{eq:C1eq}).
For $f'(i,j)$ the lines are labelled by the value of $j$. $T=0.95 T_m$.
All results are for $\epsilon_f=\epsilon$.
}
\end{minipage}
\end{figure}
\end{center}

The convergence of the thickness to its steady-state value is still described by 
a fixed-point attractor (Figure \ref{fig:attract.ef}a) but the slope of the line is closer
to 1 and so the convergence is less rapid.
Furthermore, the temperature dependence of $l_{\rm bulk}(j)$ is different (Figure \ref{fig:attract.ef}b).
$l_{\rm bulk}(j)$ is now approximately constant at a value near to $j$ over a wide range of temperature,
whereas for $\epsilon_f=0$, $l_{\rm bulk}(j)$ is approximately independent of
$j$ at low temperature with a value close to $l_{\rm min}$ (Figure \ref{fig:attract}c).
This difference is due to the greater difficulty of initiating a new stem when $\epsilon_f=\epsilon$;
the length of the stem is therefore much more likely to reach $j$.
Similarly, comparison of Figure \ref{fig:pstem.ef} with Figure \ref{fig:pstem} shows that 
the kinetic decay of $f'_{\rm bulk}(i,j)$ in the range $l_{\rm min}<i<j$ is much reduced.
As a result the steady-state and equilibrium probabilities for the outer layer are in closer
agreement than for $\epsilon_f=0$ (Figure \ref{fig:pstem.ef}c) and the deviation of the rounding
of the crystal profile from the theoretical prediction is reduced at low temperature (Figure \ref{fig:profile}c).

The slopes of the free energy profile $A_1(i|j)$ for $i<j$ and $i>j$ (Figure \ref{fig:A1}a) 
have the same magnitude when $T=0.75T_m$. 
Therefore, below this temperature the probability distribution $f'^{\rm eq}_1(i,j)$ 
is asymmetrical about $j$ with $i>j$ more likely.
This asymmetry can also be seen for $f'_{\rm bulk}(i,j)$ in Figure \ref{fig:pstem.ef}a. 
In the absence of the low $i$ shoulder in $f'_{\rm bulk}(i,j)$, which is caused
by the initiation of new stems before equilibrium in the outermost layer is reached,
this asymmetry would lead to the divergence of the thickness.
Instead, it just causes $l_{\rm bulk}$ to rise modestly.
By contrast, below $T<0.5T_m$ there is no longer any free energy barrier to the 
extension of a stem beyond the previous layer and $l_{\rm bulk}$ increases rapidly.
These effects are not seen at $\epsilon_f=0$ because of the more rapid kinetic decay of
$f'(i,j)$ as $i$ increases. 

Although using a non-zero $\epsilon_f$ makes the behaviour of the SG and ULH models
in some ways more similar, differences remain. For example, the crystal profile is 
still always rounded (Figure \ref{fig:profile}c). The differences probably stem from 
the differences in the number of growth sites in the two models.
In the ULH model the crystalline polymer is explicitly represented
and there is only two sites in any layer at which growth can occur---at either end of the crystalline 
portion of the polymer. In contrast growth can occur at any stem position 
at the growth front in the SG model.
Consequently, in the ULH model most of
the stems in the outermost layer must be longer than $l_{\rm min}$ if the 
new crystalline layer is to be stable, thus
making the rounding of the crystal profile less than for the SG model.

We can also reach this conclusion if we note that in some ways the ULH model is similar to 
the two-dimensional SG model but where the slice represents a polymer growing along the surface rather
than a cut through the crystal perpendicular to the growth face.\cite{Goldbeck94a}
For this variant of the SG model, Figure \ref{fig:profile}a would represent the dependence of the 
average stem length in the outermost layer on the distance from the growing end of
the crystalline portion of the polymer. Only near to this growing end could the stem length
be less than $l_{\rm min}$.

\section{Conclusion}

In this paper we have reexamined the mechanism of crystal thickness selection
in the Sadler-Gilmer (SG) model of polymer crystallization.
We find that, as with the ULH model that we investigated previously,\cite{Doye98b,Doye98d} 
a fixed-point attractor describes the dynamical convergence of the thickness to a 
value just larger than $l_{\rm min}$ as the crystal grows.
This convergence arises from the combined effect of two constraints on the
length of stems in a layer: it is unfavourable for a stem to be shorter than 
$l_{\rm min}$ and for a stem to overhang the edge of the previous layer.
There is also a kinetic factor which affects the stem length:
whenever a stem is longer than $l_{\rm min}$ the growth of the stem can be terminated 
by the successful initiation of a new stem.
This factor 
plays an important role in constraining the stem length at larger supercoolings 
when the free energy of crystallization is larger and the barrier to extension of a 
stem beyond the previous layer is therefore smaller.

This explanation of thickness selection differs from that given by Sadler and Gilmer.
They realized that in order to explain the temperature dependence of the growth rate 
in their model they required a `barrier' term that increased with the crystal thickness
at the steady state.\cite{Sadler84a}
They correctly identified that the rounding of the crystal
profile has just such an inhibiting effect on the growth rate.
However, they then also tried to explain the thickness selection in terms of 
the growth rate in a manner somewhat similar to the maximum growth rate approach
in the LH theory.\cite{Sadler87d}
This approach has a number of problems. 
Firstly, $G(i)$ (Equation \ref{eq:Gj}) merely characterizes the steady state. 
A maximum in that quantity does not explain how or why convergence to that
steady state is achieved.  
$G(i)$ is just the rate of incorporation of stems of length $i$ into the crystal
at the steady state. Equation (\ref{eq:GC}) trivially shows that it must have 
a maximum at the most common stem length in the crystal.
By contrast, Sadler's alternative maximum growth argument in terms of $C(l)$
has potentially more explanatory power, but it falls into the LH trap of 
incorrectly assuming that a crystal of any thickness can continue to grow at that thickness.

Secondly, it is somewhat misleading to label the effect of the rounding of the crystal profile as due to an entropic 
barrier because it is due to both the energetic and entropic 
contributions to the free energy of the possible configurations.
In particular, it ignores the important role of the free energy barrier for the extension of a 
stem beyond the edge of the previous layer.

We therefore argue that our interpretation in terms of a fixed-point attractor provides a more
coherent description of the thickness selection mechanism in the SG model.
The added advantage of this pictures is its generality---we have shown it holds for both
the SG and ULH models. Therefore, the mechanism is not reliant upon the specific features of
these models.

There is also experimental evidence
that a fixed-point attractor describes the mechanism of thickness selection
in lamellar polymer crystals. 
The steps on the lamellae that result from a change in temperature
during crystallization\cite{Bassett62,Dosiere86a} show how 
the thickness converges to the new $l^{**}$.
The characterization of the profiles of these steps by atomic-force microscopy
could enable the fixed-point attractors that underlie this behaviour to
be experimentally determined.
As well as testing our theoretical predictions, such results would 
provide important data that could help provide a firmer understanding of
the microscopic processes involved in polymer crystallization.
For example, although the mechanism of thickness selection in both the 
ULH model and the SG model is described by a fixed-point attractor, 
the form of the temperature dependence of $l_{\rm bulk}(j)$ has significant
variations depending on the model used and the parameters of the model (compare for
example, Figures \ref{fig:attract}c and \ref{fig:attract.ef}b in this paper
and Figures 6a and 13a in Ref. \onlinecite{Doye98d}).

\acknowledgements
The work of the FOM Institute is part of the research programme of
`Stichting Fundamenteel Onderzoek der Materie' (FOM)
and is supported by NWO (`Nederlandse Organisatie voor Wetenschappelijk Onderzoek'). 
JPKD acknowledges the financial support provided by the Computational Materials Science 
program of the NWO and by Emmanuel College, Cambridge.
We would like to thank James Polson and Wim de Jeu for a critical reading of the manuscript.

\appendix 
\section{Calculating C(l)}

In Ref.\ \onlinecite{Sadler87d} Sadler presented an expression for the probability that
a crystal of thickness $l$ has a squared-off profile (i.e.\ $l_n=l$ for all $n$).
To do so he assumed that only tapered configurations were possible, i.e.\ $l_n\le l_{n+1}$, 
and that the system is at equilibrium. 
The calculation requires that the ratio of the Boltzmann weight
of the squared-off configuration to the partition function, $Z$, be calculated. 
If we use the free energy of the squared-off configuration for our zero, $C(l)=1/Z$.

\begin{center}
\begin{figure}
\epsfig{figure=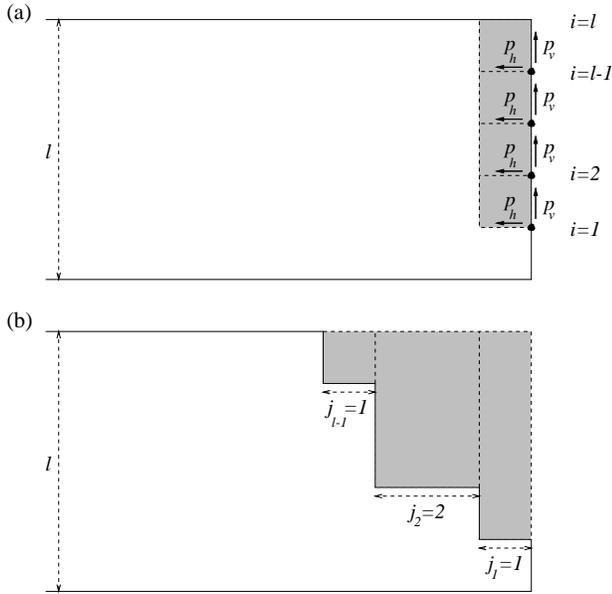,width=8.2cm}
\vglue 2mm
\begin{minipage}{8.5cm}
\caption{\label{fig:calc} 
Schematic diagram of the crystal profile to show (a) Sadler's approach and (b) our approach 
for calculating the probability of a squared-off configuration.
The shaded area is proportional area to the free energetic cost of a configuration.
}
\end{minipage}
\end{figure}
\end{center}

Sadler argues that $C(l)$ can be calculated using a random-walk analogy 
as in Figure \ref{fig:calc}a.
Each possible configuration is represented by by a walk of horizontal and vertical 
displacements from $i=1$ to $i=l$. 
To achieve a squared-off configuration $l-1$ successive vertical displacements from 
$i=1$ have to be made.
As a horizontal move would involve a loss in free energy of $-(l-i)\Delta F$,
where $\Delta F$ is the free energy of crystallization per polymer unit,
the ratio of the probabilities of horizontal and vertical moves is 
\begin{equation}
{p_h\over p_v}=\exp((l-i)\Delta F/kT).
\end{equation}
It simply follows from $p_h+p_v=1$ that
\begin{equation}
p_v={1\over 1+ \exp((l-i)\Delta F/kT)}.
\end{equation}
Therefore, the probability of $l-1$ successive vertical steps is
\begin{equation}
C(l)=\prod_{i=1}^{l-1}{1\over 1+ \exp((l-i)\Delta F/kT)}.
\end{equation}
Substituting the exponential by a ratio of rate constants (Equations
\ref{eq:kratio} and \ref{eq:kratio2}) and using $j=l-i$ gives
\begin{equation}
C(l)=\prod_{j=1}^{l-1}{1\over 1+ (k^-/k^+)^j}.
\end{equation}

However, the method of calculation is flawed. 
A proper calculation of the partition function requires that all possible 
configurations be considered.
We can characterize a general configuration by the set of integers 
$\{j_1,\dots, j_{l-1}\}$ (Figure \ref{fig:calc}b).
The free energy of a configuration with respect to the squared-off configuration
can be calculated by summing the shaded rectangles in Figure \ref{fig:calc}b; 
this gives $-\sum_{i=1}^{l-1} j_i (l-i) \Delta F$.
The partition function can then be written as
\begin{equation}
Z=\sum_{j_1,\dots,j_{l-1}=0}^\infty \exp \left(\sum_{i=1}^{l-1} j_i (l-i) \Delta F/kT\right).
\end{equation}
As an exponential of a sum is a product of exponentials
\begin{equation}
Z=\prod_{i=1}^{l-1}\sum_{j_i=0}^\infty \exp (j_i (l-i) \Delta F/kT).
\end{equation}
The sum is a geometric series so
\begin{equation}
Z=\prod_{i=1}^{l-1} {1\over 1 - \exp ((l-i) \Delta F/kT)}.
\end{equation}
Using $C(l)=1/Z$ and again substituting for the exponent and for $i$ gives
\begin{equation}
C(l)=\prod_{j=1}^{l-1} \left[ 1 - (k^-/k^+)^j \right].
\end{equation}

In this approach we can also calculate a number of other quantities.
For example, $C_1(i)=Z_1(i)/Z$, where $Z_1(i)$, the contribution to 
the partition function from configurations with an outer stem of length $i$,
is given by 
\begin{equation}
Z_1(i)=(k^-/k^+)^{l-i} \prod_{j=1}^{l-i} {1\over 1-(k^-/k^+)^j} \\
\end{equation}
for $i<l-1$ and by
$Z_1(l-1)= \left(1-k^-/k^+\right)^{-1}$ and $Z_1(l)=1$.
$\bar{l}_1$ can then be calculated using Equation (\ref{eq:lave}).

\end{multicols}

\begin{thebibliography}{10}
\bibitem{GGWrate}
An extended version of the rate equation approach can also be used to study the
  transient behaviour of the SG model.\cite{Goldbeck93a,Goldbeck89a}.
\bibitem{check}
Indeed this relationship is a good check that the integration of the rate
  equations has converged to the steady state.
\bibitem{Clmelt}
The behaviour at $T=T_m$ provides a simple physical test for the two formaulas
  for $C(l)$. Our expression correctly goes to zero, whereas Sadler's
  expression gives $2^{l-1}$.
\bibitem{Clrate}
$C(l_{\rm bulk})$ is actually an exponential function of $l_{\rm bulk}$.
  Therefore, it can be used to explain the temperature dependence of the growth
  rate. This behaviour is mainly a consequence of the temperature dependence of
  $C(l)$. As the temperature is increased it becomes increasingly hard to
  achieve a squared-off configuration (Figure \ref{fig:psquare}).
\end{thebibliography}
\end{document}